# Schwinger's Magnetic Model of Matter – Can It Help Us With Grand Unification?

Paul J. Werbos
*Program Director for Quantum, Molecular and High Performance Modeling and Simulation
Engineering Directorate, National Science Foundation[1]
Arlington, Virginia, US 22230*

**Abstract.** Many have argued that research on grand unification or local realistic physics will not be truly relevant until it makes predictions verified by experiment, different from the prediction of prior theory (the standard model). This paper proposes a new strategy (and candidate Lagrangians) for such models; that strategy in turn calls for reconsideration of Schwinger's magnetic model of matter. High priority should be given to experiments which fully confirm or deny recent scattering calculations which suggest the presence of van der Waal's effects in low energy p-p and π-π scattering, consistent with Schwinger's model and inconsistent with QCD as we know it (with a mass gap). I briefly discuss other evidence, which does not yet rule out Schwinger's theory. A recent analysis of hadron masses also seems more consistent with the Schwinger model than with QCD. Key words: Dyon, bosonization, soliton, monopole, QCD, Schwinger, stochastic quantization.  PACS – 12.60-I, 11.25.Mj, 14.80.Hv

## I. INTRODUCTION

The goal of this paper is to suggest how reconsideration of Schwinger's "magnetic model of matter" (MMM)[1] can help us overcome some of the roadblocks towards a larger goal: the development of a mathematically well-posed and unified model of how all the forces of nature work together, capable of predicting the full spectrum of empirical data from the laboratory. MMM itself is *not* such a model – but neither is anything else available to us today. Rather, MMM can be useful as a kind of tool in coping with three major roadblocks which are limiting our progress towards that larger goal. In this introduction, I will start by discussing the larger goal, and mention MMM only as it connects to some of the key subgoals.

      In all honesty – the goal of grand unification is not the only motivation here. I will argue that the most promising approach to building a *finite* unified field theory in 3+1 dimensions is to start from bosonic models which generate solitons; however, a bosonic unified theory would also have profound implications for the foundations of physics. Because of the many exact results which now exist for classical-quantum equivalence in the *bosonic* case[2,3,4], such a theory would seriously re-open the possibility of local realistic models powerful enough to address the complex empirical database of physics today. Section III-4 will discuss this further, but this paper will mainly address the issue of grand unification in 3+1 dimensions, which is certainly a challenging enough starting point.

---

[1] The views herein are not anyone's official views, but this does constitute work produced on government time.

The three greatest roadblocks to the larger goal, in my view, are:

(1) the *theoretician/experimentalist divide*, most notably the huge distance between true unified models like superstring theory and the practical phenomenological models used to make sense of the mid-to-low energy nuclear experiments which are the bread and butter of large nuclear laboratories today[5];

(2) the *physics/mathematics divide*, the difficulty of formulating nontrivial quantum field theories which are truly well-posed according to the standards of mathematicians[6,7,8] or even the more humble standards of rigorous engineers;

(3) *the mass prediction gap* (not to be confused with the mass gap[6]), the impossibility of really predicting the masses of quarks or leptons when using theories like quantum electrodynamics (QED)[9,10,11] or quantum chromodynamics (QCD)[12,11] which only become meaningful when we attach elaborate, nonphysical systems for regularization and renormalization as part of the *definition* of the theory.

The original motivation for this paper came from the theoretical side. Like the superstring people, I began by asking: "Can I come up with a well-defined quantum field theory which is *finite*, which reproduces all the tested predictions of the standard model of physics, but does not require renormalization and regularization as part of the definition of the theory?" However, unlike the superstring people, I asked: (1) can we do it without requiring additional, speculative dimensions; and (2) can we do it even without gravity, just to get started?

The biggest reason why QED *requires* renormalization is that the energy of self-repulsion of an electron will always be infinite, if we assume that the charge of an electron is all concentrated at a single point. The mass-energy predicted by a point-charge model will always be infinite, unless we adjust it in an ad hoc manner, through renormalization. Superstring theories can be finite, because they assume that the electron has a kind of nonzero radius – very small, as small as the Planck length, but that is enough. There is an easier way to achieve the same effect – by modeling the most elementary particles of nature as *solitons*[5,13], as compound systems whose charge is distributed over a finite region of space.

Of course, distributing the charge is not *sufficient* by itself to give us all that we need, but it is essentially a *necessary* condition; thus in order to get to the larger goal, this is the necessary starting point. Some physicists would worry whether there is any hope at all here; to create solitons, we need interaction terms which are not bilinear, and can any model of that sort be well-defined without renormalization? In fact, superstring theories have shown that this is possible, in principle; in any case, there is no mathematical result saying that models with third order nonlinearities cannot be well-defined without renormalization. In previous work[4], I have reviewed the extensive theoretical work and strong theorems for classical-quantum equivalence, which can provide both upper and lower bounds on energies and masses in bosonic field theories. One of the many important new opportunities ahead of us here is to exploit this equivalence, to prove that all of the Lagrangians discussed in section III do in fact yield finite well-defined theories.

Soliton models like the Skyrme model[5,13] have in fact been very popular at times in empirical nuclear physics. They have been used to confront mid-to-low energy scattering data, in regimes where QCD could not be used directly[5,14]. The argument has sometimes been that the Skyrme model provides a kind of approximation to the more fundamental model, QCD. But in my theoretical work, I was asking whether we might consider a *more fundamental* model than QCD, in which the quark itself is modeled as a soliton. In effect, I was asking whether we could overcome the *mass prediction gap* and the *physics/mathematics gap* for QED and QCD, by modeling leptons and quarks as solitons so small that we end up with the same actual empirical predictions. Instead of formulating QED and QCD as the limiting case of physically meaningless regularization models (like Pauli-Villars or fractional-dimension models)[10,11,15], we could represent them as the limiting case of a family of *finite* field theories; any member of that family, of small enough radius, would be a legitimate theory of physics, consistent with all the empirical evidence supporting the standard model.

In pursuing this approach, I kept bumping into a fundamental obstacle – that the fields which yield solitons "want to be bosonic only." I also ran across important gaps in communication between different parts of the vast continent which physics has grown into.

One gap is that many orthodox physicists assume that a theory must be *renormalizable* in a certain sense, in order to be meaningful or useful as a theory of physics. More precisely, they require that the usual sort of perturbation theory – a kind of Taylor series about the zero or vacuum state – converges. This requires that the underlying coupling constants must be less than one. However, in the known soliton models for 3+1 dimensions, and in MMM, the coupling constants are larger. This entire family of models appears to fail the test.

The answer to this gap is that *we do not really need to meet this very narrow requirement in order to be well-posed or useful*. In fact, some leaders of axiomatic quantum field theory have argued[7] that we will need to move towards nonperturbative theories and methods in order to have any hope of overcoming the *physics/mathematics gap*. Soliton models can indeed be well-defined and finite, at least as well-defined as the standard model of physics, as shown by the classic, authoritative work of Rajaraman[13]. Rajaraman still uses "Taylor series" types of perturbation, but he uses a Taylor series, in effect, about a nonzero starting point (the classical soliton solution). More creative forms of perturbation analysis have also proven essential to efficient calculation and convergence in the challenging domain of many-body practical QED calculations[16].

But how do we handle the remaining difficulty – the fact that the known well-defined soliton models are built up from *bosonic fields*? How could we reproduce the predictions of the standard model of physics, which combines both bosonic fields and fermionic fields, when we only seem to have bosonic fields available?

On occasion, major physicists have argued that we could construct well-defined fermionic soliton models by somehow assuming "classical anti-commuting fields"[17,18]. I have yet to find or construct a reasonably well-defined mathematical way of doing this which really works.

*Bosonization* has been the mainstream, most promising approach to bridging the gap between soliton models and the standard model. *Bosonization* studies how quantum field theories which are purely bosonic in nature may actually result in solitons (or point particles) which behave like fermions.

The literature on bosonization was already huge and diverse, when Makhankov et al wrote their review of the subject[5]. More recently, Vachaspati has attempted to construct a "bosonic standard model"[19] based on bosonization, but encountered certain difficulties[20] which have not yet been resolved. Makhankov et al also noted[20] that the evidence for bosonization in 3+1 dimensions is very persuasive, but still not worked out so completely as the 1+1-D case.

In order to address this roadblock and open the door to new approaches to the mathematical formulation of field theory, I have proposed[4] that we should revisit a model field theory (HtJR) independently proposed by Hasenfratz and 'tHooft[21] and by Jackiw and Rebbi[22].

*My argument here is not that HtJR is "true" or that it is more plausible, in the end, than the skyrmion model or others as a theory of elementary particles.* My argument is three-fold: (1) because of the work reviewed by Rajaraman[13], HtJR (and the simpler Georgi-Glashow model which it is based on) are extremely promising as nontrivial targets for axiomatic field theory, far more interesting than the $\varphi^4$ field theories, suitable both for nonperturbative and novel perturbative approaches; (2) because of the work of Hasenfratz, 'tHooft, Jackiw and Rebbi, and because of the properties discussed in Section III and the Appendix, we are *closer* to being able to work out the details of bosonization with this model than with other models in this class; and (3) because bosonization and axiomatic properties may actually be similar for other soliton models, in the limit as the radius goes to zero, work in this direction may be an important starting point, even if we may shift in the end to other soliton models.

I will not say more about these theoretical considerations in this Introduction. A major goal of this paper – in section III and in the Appendix – is to elaborate on these points, to review MMM, and to suggest additional model field theories for mathematical investigation. Until we begin to investigate at least some of the soliton-generating bosonic field theories capable of "bosonization," we will not have any way to try to predict or explain the masses of the most elementary particles known to physics (currently leptons and quarks).

But then there is the empirical side. Even before I started to study some of the properties of the HtJR model, I immediately noted a major difficulty in trying to use it in a "bosonic standard model": the spin-½ particles which it predicts are *dyons* – they possess both electric and magnetic charge.

Aside from what Vachaspati has already attempted[19,20], there are three obvious ways to try to deal with this problem:
(1) to construct a modified model of the quark, based on Schwinger's MMM[1], where he proposes, in effect, that the quark *is* a dyon;
(2) to *reinterpret* the topological charge in the HtJR model as a *mixture* of ordinary electrical and magnetic charge, such that the soliton only possesses electrical charge; or
(3) work out the mathematics, as proposed here, but use it only as a steppingstone to understanding the properties of alternative soliton models, such as the more difficult Skyrme model, which may or may not have fewer difficulties.

*We do not yet know* which of these three approaches will work best, in the end, in getting to the larger goal here. Thus to get to the larger goal, I would propose that the community of physicists work as vigorously as possible in *all three directions*, in parallel, without becoming overly committed to one or the other.

When I first encountered this three-fold choice, I asked myself: "What is the empirical evidence right now that the quark does *not* possess magnetic charge, as

Schwinger proposed?" I looked for definitive papers, similar to the classic three tests of general relativity versus Newtonian gravity, which would rule out Schwinger's model or define what is needed to rule it out. To my great surprise, after a very thorough search, I found very little work addressing this head-to-head comparison. So far as I know, Sawada of Japan is the only person who has gone directly to the empirical data to try to find out what they say about the QCD-versus-Schwinger issue. So far as I know, Feinberg is the only living person who has directly addressed the theoretical implications of Sawada's calculations. To my greatest surprise, these preliminary indications appear to support Schwinger's model over QCD as we now understand it (with a mass gap). Today's best understanding of atomic nuclei (Section II-4) also suggests that the strong nuclear force may be longer-range than we would have thought form QCD.

*I am not suggesting that this preliminary evidence disproves QCD*. On the contrary, I am suggesting that we need to change this situation, by getting better evidence. Because of the great importance and great difficulty of the larger goal here, I am also proposing that we should at least work hard to do full justice to the *possibility* that approach (1) might work out in the end.

Since I have started getting a bit deeper into these issues (see Appendix), I see more and more hope for approach (2). Perhaps this will reduce the real need for approach (2). Nevertheless, I do hope that someone will be able to follow up on approach (1) as well, at least by conducting more decisive experiments. Because I am really coming at this from the mathematical side, and because the empirical literature here is huge, my review in section II is incomplete; however, I hope it will provide at least a starting point for the empirical nuclear (or electromagnetic) physicist, by giving some sense of what is available in the more complete and detailed papers I cite. For more details, of course, it would be easy enough to look at the electronic versions of these papers at arXiv.org and at the APS web site.

## II. THE EMPIRICAL SITUATION FOR MMM VS QCD

A final theory of strong interactions should allow us to predict at least:
(1) High energy scattering data – the asymptotic limit as energy goes to infinity;
(2) Low and medium energy scattering data – in the entire data set, including correlation effects;
(3) The spectrum of masses of the hadrons and quarks, and the internal binding which causes these particles to exist at all;
(4) Many-body effects involving many hadrons.

The original quark model – an SU(3) theory assuming symmetry between u, d and s quarks – was motivated by an effort to explain mass spectra[23]. QCD emerged later as one of several possible modifications to that model[12].

QCD has had excellent success in quantitative predictions of several high-energy scattering measurements. However, such predictions are essentially the first order terms in a Taylor series expansion in (1/E). We still have many degrees of freedom that we can change, without changing these confirmed predictions. So far as I know, there is no evidence as yet that these experiments rule out the possibility of magnetic charge, or even that magnetic charge could be used as an alternative to color in replicating the same predictions.

A few years ago, the empirical situation seemed to be as follows. QCD was successful in high energy scattering predictions. It could also predict the masses and types of hadrons reasonably well, by adding up the masses of their *constituent quarks*; the quark masses, in turn, were chosen so as to make the hadron masses work out. With scattering at low to medium energy, the predictions of QCD were so hard to calculate that empirical researchers relied mainly on "phenomenological models," which could commonly have 20-50% errors in prediction. One of these "phenomenological models," the Skyrme model[5] was found to be the "approximation" of QCD in the case where N=3 is approximated as N=∞; however, empirical tests of the Skyrme model did not tell us much about the accuracy of QCD, when errors on all sides were on the order of 20%. QCD did make some interesting many-body predictions, for quark-gluon plasmas, which had yet to be confirmed.

Empirical data from CERN has confirmed fundamental discrepancies in the conventional form of the constituent quark approach to hadron masses[24]. At Lawrence Livermore (and then later, in retirement), Malcolm MacGregor has shown how to resurrect the constituent quark approach[25,26], with much higher accuracy than earlier studies, but only by assuming that the quarks in baryons have different masses from those in mesons like pions and kaons. His results also show that α, the fine structure constant for *electromagnetism*, seems to dominate all the masses of hadrons, as well as the muon and τ lepton. Recent calculations by Sawada[27,28], using new recent high-precision empirical data on p-p scattering[29,30,31] and π-π scattering[32] at low energies, support earlier indications[33-36] of a long-range Vanderwaals component in strong nuclear forces.  Earlier detailed calculations by Feinberg and Sucher[37] showed that such long-range effects cannot be reconciled with QCD, unless we make fundamental changes in our present understanding of QCD; in particular, we must assume a violation of the "mass gap" assumption, which is fundamental to our present understanding[6]. If the "mass gap" can be violated, there is no reason in principle why we could not convert protons and neutrons completely to energy – a "third generation of nuclear energy" orders of magnitude beyond fission and fusion; thus if QCD and Sawada's calculations *both* happen to be correct, it would be very important to know this. New experiments would be crucial to finding this out, if it should be true.

The remainder of this section reviews the four main areas of empirical data that bear on the QCD versus MMM choice. While I try to do my best to be an honest "devil's advocate" for magnetic charge, please remember that this does not reflect a commitment to the MMM choice; rather, it reflects a belief that we need to do full justice to "the devil's view" in order to work effectively towards a more complete understanding.

## II.1. Color and The Usual High-Energy Data

Again, QCD has performed well in predicting many high-energy measurements. The present article on QCD in Wikipedia does a nice job of summarizing the mainstream empirical case for the present form of QCD. Their empirical story is consistent with what I have said.

There is one point where I question the usual *interpretations* of these experiments. I do not believe that they have confirmed the existence of "color." It is common, for example, to see papers which "confirm" color by comparing today's QCD versus today's QCD with color removed. Indeed, when color is simply removed, the binding forces are weakened, and the resulting theory is invalid. However, as Schwinger[1] and others pointed out long ago, the magnetic forces which he predicted from his alternative approach are quite enough to hold the proton together without any gluons at all. This does not *disprove* the existence of gluons, but it does imply that we should seriously study the possibility that they do not exist after all. This is just one example of the fact that there are many ways we could replicate the *first order* term in a Taylor series in 1/E.

The idea of color originated from the search for a new SU(3) symmetry – in order to create a renormalizable theory – when the old u/d/s symmetry became unworkable. But the Introduction has already summarized why we should not restrict ourselves to theories which are renormalizable in the traditional sense. It is better to be finite and mathematically well-defined *without* renormalization.

Color has also been important in rationalizing how we could have two up quarks in the same hadron. But when quarks are modeled as composite particles (solitons) with unobserved internal degrees of freedom (as in some of the known solutions[21,22] of the HtJR system), we may not need such a rationalization. The HtJR system still makes heavy use of symmetries, but does not require color.

Even at high energy, collisions have been observed which clearly do not involve the exchange of color. It is claimed that these can be explained as the result of "pomerons," compound bodies which are predicted to exist in *some form* by present-day QCD. However, it may be easier and more natural simply to do without color at all, in addressing such phenomena.

Reviewers have asked: "What about the weak coupling that we observe between quarks when the energy levels rise? Is that consistent with the strong coupling constants that Schwinger has calculated here?" Actually, at high energy, we observe a combination of a weak effect coupling in some respects, but strong confinement, in current experiments. Even in a finite field theory – which is well-defined without renormalization – the effective coupling constant at high energy is not the same as the bare coupling constant, in the general case; for example, two have been calculated, and are quite different, for the simple sine-Gordon model[62]. For the moment, the effective coupling constant at high energies in various formulations of the Schwinger model is unknown. To test that model, it will be very important to make that calculation, starting from specific candidate Lagrangians, like those proposed in section III. For the moment, however, comparisons at high-energy are not available between QCD and the Schwinger model; the only direct comparisons now available are those for mass spectra and low energy scattering, as well be discussed next.

## II. 2. Mass Spectra and Existence of the Hadrons

The quark model owes its birth to efforts to explain and systematize the masses of hadrons[23]. In updating that model, it is logical to start from the most comprehensive statement of the empirical status of hadron masses – the work of Malcolm MacGregor[25,26].

MacGregor's work for Lawrence Livermore in systematizing hadron masses and lifetimes was published in a long series of journal articles over many years. I will not cite those articles directly, because his new book[25] in press reviews and integrates that work far better than I can. His recent paper at arxiv.org[26] gives a very lucid summary. In general, his work is similar in flavor to the early work on atomic spectra and fine structure which laid the foundation for quantum mechanics in the early twentieth century.

MacGregor has found that the mass spectrum of hadrons is neatly organized into two groups – a "muon mass tree" and a "pion mass tree." The constituent quark approach works very well *within* each of these mass tress, but different building-block masses apply to different trees. This is very problematic, if one assumes that different up quarks all have the same mass (because variations in color and spin direction do not change mass). But if different up quarks possess different magnetic charge, it is exactly what one would expect!

MacGregor also shows how the masses of the quarks themselves follow an orderly pattern, which cries out for explanation. The pattern strongly involves $\alpha$, which strongly suggests that *electromagnetic* forces are somehow the main thing holding together the quarks and defining their properties!

Looking more carefully – MacGregor points out that the predicted and observed masses of the $W^{\pm}$ and Z mesons already fit an "$\alpha$" based power law. Thus electroweak theory (EWT) already provides a partial possible explanation for what is going on here. But how can we use these extended electromagnetic types of forces and effects to explain the generations of quarks and leptons, which also fit an $\alpha$ scaling rule? The obvious approach is to use these kinds of forces in our model of the quarks themselves, as we will discuss in Section III.

David Akers of Lockheed-Martin has worked closely with MacGregor in recent extensions of this work. So far as I know, Akers was the first to write down a Lagrangian[38] which tries to express the initial version of Schwinger's "magnetic model of matter"[1]. Akers has also shown how assuming magnetic charge improves the empirical predictive power of the MacGregor approach[39].

Paolo Palazzi of CERN has verified the basic mass systematics of MacGregor, and provided more refinement. His analysis demonstrates a kind of fine structure to the patterns, evidence which strengthens the conclusion that there is a real pattern here crying out for a more fundamental explanation[64-65]. At a phenomenological level, the hadron spectrum appears to fit a lattice model, in which the constituents are the stable leptons, consistent with the electromagnetic model of hadrons proposed by Barut[67]. In Palazzi's initial work, the same of lattice which fits the masses of atomic nuclei appeared to fit here as well[66]; however, current work points to a modified lattice[68].

## II-3. Long-Range Effects in Low Energy Scattering

### II-3a. Overview of the Work of Sawada

In 1969, when Julian Schwinger proposed his magnetic model of matter, the scientific method demanded that we try to answer the question: "Which is right, Gellman's model or Schwinger's or something else? How can we design and perform *decisive empirical tests* to tell us which is true?"

For a variety of reasons – some legitimate and some circumstantial – few nuclear physicists rose to this challenge.

Some nuclear physicists assumed, reflexively, that the obvious way to test Schwinger's theory would be to search for isolated magnetic monopoles or dyons. Such searches have not attained any success as yet[40] – but neither have the searches for isolated quarks or other isolated elementary particles with fractional electric charge. Schwinger argued[1] that we should not expect to find isolated dyons, for theoretical reasons.

To test Schwinger's theory, using data available to mainstream empirical nuclear physics, Sawada reasoned as follows: we do not have access to a source of magnetic monopoles or dyons, but we do have the ability to detect the large magnetic dipole effects which are predicted by this theory. Schwinger's theory implies large Vanderwaals effects, much larger than what would be predicted by a charge-free gluon theory. At present, the most precise and complete data that we have available, to compare the two theories, is the standard s-wave phase shift data from low-energy p-p scattering[29-31,36]. The outcome of his work: A straightforward, fully documented, unbiased analysis of this data strongly favors the existence of such large Vanderwaals effects[27,36]. Analysis of $\pi$-$\pi$ scattering[28] fits the same conclusion.

Because these results appear to contradict our present understanding of QCD, the proper response for a scientific skeptic is to demand new experiments, to test predictions of the Vanderwaals hypothesis more flagrantly different and unusual from what would be expected from today's QCD. One way or another, such experiments are a key part of what we need to do in order to understand what is going on here. If these experiments do not support the Vanderwaals hypothesis, then the foundations of present-day QCD (including the mass gap hypothesis) would be far more secure than at present. If they do support it, then they are a crucial source of data in revising our understanding. The new results discussed in section II-2 create enough reasonable suspicion of magnetic charge to warrant such experiments.

Sawada has proposed new experiments[41] on p-p scattering with a high degree of precision in angular moments, at 40-60 Mev, enough to test for the interference patterns predicted by the Vanderwaals hypothesis. Because they could be done at a single laboratory (without piecing data together from multiple laboratories, as in the usual standard data compendia), the comparisons could be both robust and subject to testing in competing laboratories. These predictions are so different from the usual expectations that they should have the highest priority, all things being equal. His proposals for high-precision e-e scattering experiments[28] and n-Pb experiments[42] also warrant follow-on. Higher resolution, high precision scattering data will be an important dataset for physics, no matter what theory ultimately prevails.

**II-3.b. Additional Background on p-p Scattering**

Modern theoretical textbooks sometimes say: "We once thought that pi mesons, as proposed by Yukawa, were the main mediators of the strong nuclear force, just as photons are the mediator of electromagnetic attraction and repulsion. We now know that this is false, and that gluons are the dominant mediator."

It is easy to see that this is not strictly true, by reviewing old but well-established curves on nucleon-nucleon scattering[43]. When low-energy protons are slightly deflected by stationary neutrons, it is well known that they commonly exchange charges. The proton often changes velocity very little, but changes into a neutron. Such interactions are obviously mediated by a *charged* particle (mainly $\pi^{\pm}$), not by a neutral gluon.

Most calculations in practical device electronics still rely on the concept of an electron moving in a potential well, $V(\underline{x})$, or on models of two-electron potentials $V(\underline{x}^{[i]}, \underline{x}^{[j]})$. In a similar manner, practical empirical nuclear physics at "low" or "medium" energy (which includes the kind of "low" energies observed in nuclear bombs) still makes very heavy use of two-particle effective potentials. These potentials offer an effective if approximate device for summarizing complex scattering data, and are also used as a tool to approximate calculations for many-body effects like fission and fusion.

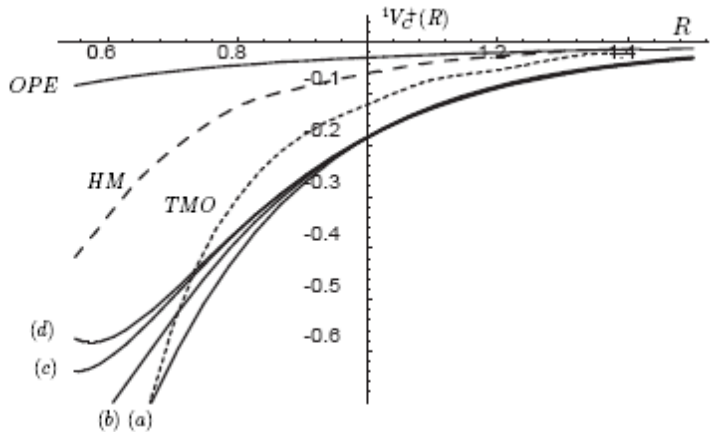

The figure above – taken from Sawada[27] – shows the effective potential one would estimate today empirically (curves (a)-(d)) assuming one-boson exchange potentials, versus the predictions one would derive from one-pion exchange (OPE), one and two pion exchanges calculated by static perturbation theory (TMO), and TMO corrected for full recoils (HM). They show a huge unexplained gap along the y axis, where the radius R is in the neighborhood of 1.

This discrepancy was known as early as the 1970's, when the Wisconsin data became widely available. In those days, an artificial one-pion+\sigma model was created, to try to generate "predictions" closer to the observed data. Imai and Nishimura did new, more precise experiments (the "Kyoto data") at 5, 7 and 8 Mev. At the end of their classic paper on this experiment[44], they firmly concluded that the one-pion+\sigma model could not explain the continuing discrepancies.

Of course, if your goal is simply to fit the empirical data, without any theoretical model, it does become possible to achieve some accuracy simply by including dozens upon dozens of parameters. That has been accomplished by many groups. (See the lengthy reviews in Sawada[35,36].) This can be very useful, in the absence of a more complete empirical theory – but it does not explain *why* the discrepancies exist.

**II-3.c. Other Work on Nuclear Vanderwaals Effects**

During the 1970's, when QCD was not so well understood as it is today, there were considerable efforts to try to explain the long-range discrepancies on the basis of Vanderwaals effects *within* the context of orthodox QCD itself. So far as we can tell, this line of research ended after the publication of a review paper by Feinberg and Sucher[37] in

1979. That paper mainly attempted to analyze what QCD actually predicts or allows for such long-range effects, though it used some empirical data in setting such limits.

Using cautious language, they stated (page 1725): "Since present formulations of QCD have not, to our knowledge, been shown to be equivalent to one in which such axioms hold [violation of mass gap etc.], we cannot draw a firm conclusion regarding the incompatibility of long-range forces with QCD. Since there apparently is a mass gap in the physical spectrum of hadrons, the discovery of a strong long-range force between hadrons might have a significant bearing on the field theoretic foundations of QCD." On page 1733, they express some doubts regarding Sawada's papers at that time, but their only specific argument is that we would need more confirmation for results so much at odds with what QCD says we should expect. In fact, the analysis of more recent, more accurate and widely available data does provide considerable extra confirmation – enough at least to warrant performing more decisive experiments.

Feinberg and Sucher also made fascinating observations about other Japanese work which suggested relatively long-range nuclear effects within bound systems like molecules. They thoroughly discredited those suggestions, within the context of QCD, by arguing (page 1733) that nuclear systems do not possess the additional degrees of freedom which justify the kinds of calculations used in normal QED calculations for molecules. Yet the soliton field-based models we propose may well reintroduce such degrees of freedom, and reopen these issues.

## II-4. Many-Hadron Effects

Many-hadron effects may be extremely important to future technology, but the research strategy proposed here does not address them directly. As with the history of Quantum Electrodynamics (QED), we need a coherent push to understand the single-hadron cases at the start, in order to better prepare for that future.

QCD has made interesting predictions for many-body effects such as quark-gluon plasmas. But these have not been confirmed, and are not part of our story here. Some researchers at the Institute for High Energy Physics (IHEP) at Protvino in Russia have listed this issue, and a handful of others, as evidence for their view that electromagnetic effects, instead of color, explain most of what they observe in practical, empirical work[45]. I have not yet studied their claims in detail.

The main empirical evidence available today on many-body nuclear effects is the extensive evidence available on the properties of the >2,000 varieties of atomic nucleus, across elements and isotopes. Normal Cook[62] has recently reviewed the thirty or so phenomenological models now used to predict or explain these properties – cluster models, liquid drop models and more traditional quantum models with a Hartree-Fock flavor similar to what is used in quantum electronics. He notes that each of these types of models has had great success in predicting *some* of these properties, but not all. He provides a new model (closer to the quantum models in underlying properties) which is able to predict the entire range of properties from a single, unified model. I thank Paolo Palazzi of CERN for making me aware of this model – and its great success in predicting the masses of nuclei. In his analysis, Cook explains how there appears to be a very serious possibility (from the empirical data) that the strong nuclear force might indeed be longer-range than we believe today based on QCD. In other words, there is additional evidence here that we should be revisiting the issues discussed in section II-3.

Many of the details of the strong nuclear force may seem academic or higher-order in nature at present. But if the analogy to QED holds, we may be surprised in the future, if we develop enough fundamental understanding to harness many-body effects. Early work on coherence in electromagnetism turned out to be extremely important, later, to the understanding of multibody QED. Issues which seemed to be of mere academic interest in the early twentieth century later became extremely important to technology, *after* they were better understood and many-body applications were developed. Tools like lasers are making it possible to probe and fine-tune electromagnetic effects, and exploit new degrees of freedom, far beyond what seemed possible even in the 1960's. Now that atom lasers and X-ray lasers have also been developed, it is quite possible that similar developments may be possible for strong nuclear forces as well, after we develop the prerequisites. If any of the unified models suggested in section III have any degree of truth, then the development of vector meson lasers[46] could open up many new directions. Many of the mathematical tools developed for many-body QED[2,3,4] work best for *bosonic* fields; thus the new models proposed here may be useful in making it possible to use them in the nuclear case.

Schwinger[47] and Jackiw[18] both hoped to apply some of these methods directly to fermionic fields as well; however, computational work in applied QED[48] has made it clear that methods like the Wigner representation require far more complexity for fermions like electrons than the same representation as applied to electromagnetism[6,7]. Notions like "anticommutative classical fields" are not so easy to operationalize in a rigorous, practical way.

Because our knowledge of many-body nuclear physics is still so incomplete, it may be prudent to consider earth orbit as a venue for future experiments which truly probe our greatest areas of uncertainty. If new technologies do turn out to be possible here, it is possible that they may require caution, as is the case for fission and fusion; however, just as nuclear energy turned out to be a key element in addressing energy challenges we did not really anticipate seventy years ago, new nuclear technologies may give us a chance to address future challenges. If there is any hope at all of the human species reaching out to the stars, it would probably require new types of technologies – and the best for that lies in firmly addressing our greatest uncertainties about the strongest type of forces that we know of, the nuclear strong forces.

## III. OPTIONS FOR IMPLEMENTING MAGNETIC CHARGE IN FUNDAMENTAL PHYSICS

In his magnetic model of matter[1], Julian Schwinger, one of the founders of quantum field theory, was proposing a fundamental theory of fields, to be part of a unified model of physics. But he did not actually write out a complete Lagrangian, and he did not consider the possibility of exploiting bosonization.

No one on earth is ready yet to write down a reliable translation of Schwinger's idea into a complete and unified Lagrangian. To achieve this, we must first understand the mathematical properties of intermediate models along the way. Therefore, this section will first review the basic ideas of Schwinger and of magnetic charge, and then proceed to describe a graded series of model field theories. It is hoped that this series of graded field theories will provide a pathway either to complete realization of Schwinger's program, or – depending on how the mathematics and empirical data work out – a realization of approaches (2) or (3) described in the Introduction.

The three model field theories are:
(1) The *steppingstone* model – modeling the quark as a solution of the Hasenfratz-tHooft-Jackiw-Rebbi[21,22] (HtJR) system;
(2) The *alternative standard* model – modeling the quark as a solution to the HtJR system *modified* by replacing the $A_\mu^a$ field with the W and Z fields of electroweak theory (EWT).
(3) A *unified* model

Schwinger's classic paper on a magnetic model of matter[1] actually refers to *two* ideas for how to inject magnetic charge into theoretical physics. The simplest idea is to model the quarks as having ±1/3 and ±2/3 magnetic charge, as well as electrical charge. That simple change, by itself, would already be enough to fit the empirical results from Sawada, as I discussed in section II. Schwinger does note that that simple model is not enough to explain CP violations, which have been studied extensively in the recent "Babar" experiments at CERN. But QCD does not explain those violations either. The only real known problem with Schwinger's initial idea, as a way of modifying QCD, is that the coupling strengths in Schwinger's model are too strong to allow a renormalizable model. The *steppingstone model* is simply a way of implementing Schwinger's initial ideas in a mathematically well-defined Lagrangian model, which offers a new starting point for axiomatic quantum field theory with a new approach. (It would be easy enough to add color as an extra degree of freedom into that model, but it is easier to begin without it, and it may indeed be unnecessary, so far as we now know.)

On the other hand, I agree with Schwinger that these initial ideas are only a steppingstone towards a more unified model of physics. For example they can be made to fit the new results on mass spectra, but they cannot really *explain* the beautiful patterns observed by MacGregor on "the power of alpha"[25,26], and they cannot explain why we see three generations of quarks. They also leave open questions about the nature of the electron and the neutrino, and perhaps about symmetry breaking. The alternative standard model and unified model below could provide a way to incorporate some of Schwinger's later ideas and to address these larger questions. Because they are further in the future than the steppingstone model, we cannot be sure we are seeing the details so clearly for those options. Probably the Lagrangians we propose as a starting point will require some debugging, after the essential work is done to better understand the soliton solutions of the HtJR models and similar systems.

It is entertaining to consider that the Lagrangian in section III-4 (after some debugging) might be enough to explain all of the phenomena we have ever observed in physics as yet. Perhaps the greatest value of such a computable, unifying model is that it allows us to predict, design or imagine new physical phenomena and alternative models, to probe even further … and generate new paradoxes, uncovering possibilities which go even further. But it will be enough of a task for us for now to try to get that far.

## III-1. From Magnetic Charge to Schwinger: A Brief Review

First, we summarize a few highlights from Milton's excellent review[40] of earlier work on magnetic charge. Benjamin Franklin was perhaps the first to propose the idea of magnetic charge explicitly. Poincare and Thomson, the discoverer of the electron, observed that the concept of magnetic charge results in a significant simplification of Maxwell's Laws. With the addition of magnetic charge, $\rho_m$, the classic version of Maxwell's Laws for the vacuum becomes:

$$\nabla \cdot \underline{E} = 4\pi\rho_e \qquad (1)$$

$$\nabla \cdot \underline{B} = 4\pi\rho_m \qquad (2)$$

$$\nabla \times \underline{B} = \frac{1}{c}\frac{\partial}{\partial t}\underline{E} + \frac{4\pi}{c}\underline{j}_e \qquad (3)$$

$$-\nabla \times \underline{E} = \frac{1}{c}\frac{\partial}{\partial t}\underline{B} + \frac{4\pi}{c}\underline{j}_m \qquad (4)$$

This system of equations is unchanged by any rotation between electricity and magnetism, i.e.:

$$\underline{E} \rightarrow \underline{E} \cos \theta + \underline{B} \sin \theta \qquad (5)$$

$$\underline{B} \rightarrow \underline{B} \cos \theta - \underline{E} \sin \theta, \qquad (6)$$

where θ is any real angle and the current vectors $\underline{j}$ are also rotated. There was no empirical evidence for $\rho_m \neq 0$ at that time, but no theoretical difficulties either.

In the first version of quantum mechanics, the electromagnetic field was represented by the field $A_\mu$ instead of by $\underline{E}$ and $\underline{B}$. Electrons and protons were described by the Dirac equation. Dirac[49] argued that magnetic charge could still be included in the theory, in principle, but *only if* the magnetic charge g and the electric charge e met a certain joint quantization condition.

Modern electronics is based on quantum electrodynamics (QED)[9]. Schwinger, Feynman and Tomonaga shared the Nobel Prize for discovering QED and the canonical version of quantum field theory (QFT). The canonical version is still in wide use today, and makes contact very directly with empirical reality[10]. The newer functional integral version of QFT[11] is essentially a synthesis of two competing formalisms which competed in 1970 – the "path integral" approach of Feynman and the "source theory" approach of Schwinger[17].

In 1966, Schwinger revisited the possibility of magnetic charge from the viewpoint of relatively canonical QED[50]. In 1968 he probed the issue more deeply from the viewpoint of source theory[51]. In his classic 1969 paper on a "magnetic model of matter"[1], he argued that:

(1) The very existence of magnetic charge would force a new quantization condition, which we need in order to explain a mystery of central importance: why electric charge exists only in exact integer multiples of e, in all free particles and systems;

(2) The correct quantization condition is:

$$(e_1 g_2 - e_2 g_1)/\hbar c = \nu, \qquad (7)$$

where ν must be an integer, where the condition must be met for *all* pairs of particles 1 and 2 in the universe, where $e_1$ and $e_2$ are the electrical charges of two particles, and where $g_1$ and $g_2$ are their magnetic charges. From this condition, he deduced that the unit of magnetic charge $g_0$ obeys:

$$g_0^2/\hbar c \approx 4(137) \qquad (8)$$

He immediately concluded that forces between magnetic charges are superstrong in comparison with the strong nuclear forces for which coupling constants are about 10.
    Because these magnetic effects are so strong, we would not expect to see free particles – elementary or compound – with nonzero magnetic charge, except perhaps in temporary events requiring very high energy. From this Schwinger deduced that the electric charge e which we see in free particles is not the unit of electric charge $e_0$, but $3e_0$!! He predicted that the elementary components of hadrons would have electric charge of

$$\pm n e_0 = \pm (n/3) e \qquad (9)$$

and *magnetic charge likewise*! For the constituents of hadrons, he proposed a model in which n=1 and n=2 are the two allowed states. The steppingstone model of Section III-2 is proposed as a way to implement this idea; more precisely, the idea is to choose parameters such that the stable soliton states fit n=1 and n=2, as Schwinger proposed.
    This paper by Schwinger was the first to use the term "dyon" as a name for particles which possess both electrical *and* magnetic charge.
    Equation 7 from source theory was not universally accented. Many researchers cite an earlier paper Zwanziger[52] which derived an alternative quantization condition from more traditional QFT approaches. In footnote 4 of Schwinger's paper[1], he claims that Zwanziger is off by a factor of 4, but leaves it to the reader to compare the two papers. Traditional quantum thinking requires very subtle judgments here, and in the case of spin as well[53].
    Fortunately, if we multiply or divide ν in equation 7 by a factor of 4, we still arrive at equation 9 and the same general model. In the steppingstone model, the quantization of magnetic charge results directly from the use of *topological charge* to model it; however, the quantization of electric charge[13] still is based on Schwinger's quantization rules. In the more fully unified model (section III-4), the quantization of electric charge and magnetic charge would *both* fall out from topological charge; in that case, the usual quantization rule and the parameters in that rule would essentially fall out as emergent predictions of the model rather than axioms of nature.

## III-2. The Steppingstone Model

The steppingstone model postulates that the quark is a stable soliton state of the HtJR system, which is specified as equation 3 of Jackiw and Rebbi[22]:

$$\mathcal{L} = -\tfrac{1}{4}F_a{}^{\mu\nu}F_{a\mu\nu} + \tfrac{1}{2}(D^\mu\Phi)_a(D_\mu\Phi)_a - (1/e^2)V_1(e^2\Phi^2) + (D^\mu U)^\dagger(D_\mu U) - (1/e^2)V_2(e^2U^\dagger U, e^2\Phi^2);$$
$$F_a{}^{\mu\nu} = \partial^\mu A_a{}^\nu - \partial^\nu A_a{}^\mu + e\epsilon_{abc}A_b{}^\mu A_c{}^\nu;$$
$$(D^\mu\Phi)_a = \partial^\mu\Phi_a + e\epsilon_{abc}A_b{}^\mu\Phi_c;$$
$$(D^\mu U) = \partial^\mu U - ie(\tau^a/2)A_a{}^\mu U;$$
$$\Phi^2 = \Phi_a\Phi_a, \quad a = 1, 2, 3;$$
$$V_1(\Phi^2) = (\lambda^2/2)[(\mu^2/\lambda^2) - \Phi^2]^2;$$
$$V_2(U^\dagger U, \Phi^2) = m^2 U^\dagger U + g^2(U^\dagger U)^2 - h^2 U^\dagger U[(\mu^2/\lambda^2) - \Phi^2].$$

(10)

Notice that there are six parameters here – e, m, g, h, μ and λ. All three fields -- Φ, A and U – are bosonic fields. Nevertheless, Hasenfratz and 'tHooft[21] and Jackiw and Rebbi[22] both showed that this system yields solutions which have spin one-half, and possess both magnetic and electric charge. For example, Jackiw and Rebbi reported solutions of the form:

$$A_a{}^0 = 0 \tag{11}$$
$$\Phi_a = \hat{r}^a(\varphi(r)) \tag{12}$$
$$\vec{A}_a = \hat{\eta}_a(A(r) + 1/er), \quad a = 1,2 \tag{13}$$
$$\vec{A}_3 = \vec{A}_D(\vec{r}) \tag{14}$$
$$U = u(r)\operatorname{Re}xp(-i\vec{\alpha}\cdot\vec{\tau}/2)s, \tag{15}$$

where R is an isorotation matrix. Analyzing these solutions in collective coordinates, he shows that the spin of the overall solution has somehow become equal to the isospin; a spin of ½ and fermionic statistics have resulted as an *emergent outcome* of underlying dynamics which are purely bosonic in nature!

Goldhaber later re-analyzed and confirmed this effect[53]. Schwinger[1] explained long ago why such dyons, *within a composite system* (like a hadron), can possess composite charges which we interpret as ±1/3 or ±2/3.

The initial idea here is that an up-flavored quark, for example, may exist in either a +2/3 magnetic charge form or a -2/3 magnetic charge form. But magnetic charge in total must be zero, for any composite body, whether it is a baryon or a meson. This implies that we only see one combination of fractional magnetic charges within a baryon, and another one which may be different within a meson (depending on the type of meson). When we first noticed this prediction, it sounded highly implausible, since it would imply different masses for constituent quarks in different compound bodies; however, that is exactly what MacGregor has observed.

Why pick this particular system, when there are other possible dyon models out there?

The main reason is that this is the only system in three spatial dimensions which is now known to generate fermion compound particles from bosonic, underlying fields. We need a spin-half particle to model the quark, and we need the bosonic fields and soliton mechanism in order to achieve a *finite* (finite-mass without renormalization) field theory. ***Only a theory which actually predicts finite masses can predict quark masses; renormalizable theories which insert masses cannot do so.***

Another reason is that we would not actually use (as yet) the detailed predictions for the internal structure of the soliton. The radius, r, of the predicted soliton depends on the six parameters. For a finite theory, r must be finite – but we have no empirical data as yet to estimate how small r is. For all we know, r might be as small as what superstring theory assumes – about a Planck length. It is premature to worry about alternative possibilities that we could not distinguish from each other.

Palazzi and others have proposed that we can explain the emerging spectrum of quark and lepton masses and lifetimes by modeling these "elementary particles" as bound states with radiative decay modes, analogous to how we explained ordinary atomic spectra in the early twentieth century. My suggestion here is actually compatible with that program. The fermionic bound states in the work of HtJR are actually functions of six spatial coordinates, similar in form to the old calculations for the hydrogen atom. In the nuclear case, programs like those of Palazzi require strong coupling constants, inconsistent with the more traditional types of models; here I am pointing out a mathematical pathway which can allow their program to go forward, in a mathematically well-defined (and falsifiable) family of theories.

What we do need to use is the spectrum of states. We do need to know sets of values of the six parameters (e, m, g, h, $\mu$, $\lambda$) which yield stable soliton solutions of magnetic and electric charge $\pm 1/3$ and $\pm 2/3$, with spin ½, which can fit (at least) the up and down quarks. We can expect that this will work, from the previous work on this system, but more complete analysis and verification is needed.

Furthermore, we would really want to find a parameterized family of such parameter values – i.e. three sets of functions e(r), m(r), g(r), h(r), $\mu$(r) and $\lambda$(r) for the three generations of quarks – such that the limits as r goes to zero match what we observe in nuclear data. This may sound similar to regularization, but each value of r along the path corresponds to a well-defined mathematical field theory. Any small enough nonzero value of r would correspond to a well-defined theory which fits the known data.

Hasenfratz and 'tHooft note that their work focuses on parameter values which yield a pure bosonic state as the (stable) state of lowest energy. Here, we must choose a family of parameter values in the sector studied by Jackiw and Rebbi, where the (stable bound) state of lowest energy is a fermion. In the limit as r goes to zero, but the physical mass goes to an observed quark or lepton mass, I would expect that the mass of the bare bosons would go to infinity.

This model does not attempt to explain CP violations, as in the more complex model suggested in the end by Schwinger[1]. For simplicity, we propose initial work on something more like his initial concept (discussed in the same paper). Because parity violations are far more visible in weak interactions than in strong interactions, and superweak interactions would seem to be more aligned with weak interactions, we propose that CP violations should be addressed instead either in modified EWT or in a later stage of the unified model for the future.

There are three key requirements for new theoretical work here.

One is to prove that the HtJR system is indeed an axiomatically well-defined system. At present, the $\varphi^4$ model is the only nontrivial field theory close to being mathematically well-defined (though the details are controversial), using conventional approaches. But the kinds of mathematical approaches used by Rajaraman[13] offer a different pathway to proving existence of a field theory. Methods of classical-quantum correspondence offer lower and upper bounds to masses[4,54], which should make it reasonably straightforward to prove finiteness.

Another key requirement is a more complete characterization of the complete spectrum of solutions – including their spin and stability properties. The HtJR system is an extension of the Georgi-Glashow system (which had $\Phi$ and A but not U), whose solutions have been studied in enormous detail. Similar theoretical work is needed here.

The discussion of spin in this system is far more convincing than what we have seen in other parts of the vast literature on "bosonization" in 3+1 dimensions. When a compound particle (or soliton) of unmeasurably small radius has a spin ½, it is easy to assume that it *must* obey the usual fermion statistics, because of the usual spin-statistics theorem[8]; however, that remains to be proven (see the Appendix), and the details are of great physical significance. The possibilities for electric and magnetic charge should fall out in a direct way from that approach.

This last point has some deep implications. Collective coordinates have been a central part of the analysis of the HtJR system so far. If the approach in the Appendix should break down, then the classical solution over six dimensions (like the "two-body Schrodinger equation") may provide an essential starting point, even for a WKB approach to defining the theory. The convergence to $\Psi(R)\psi(\underline{r})$ is a mathematical device for expressing the idea that fermions may be like chaotic classical (bosonic) objects[53] at a fine scale of distance ($\underline{r}$), even though all we observe as yet is $\Psi(R)$.

Likewise, evaluations of stability are essential here. The work on stability for skyrmions[5] provides at least some starting point, more rigorous than what we have seen applied as yet to the HtJR system. Yet there is room for a more careful, systematic approach to stability analysis[55].

Finally, of course, the straightforward matching of the new quark properties to what we observe in the data is the most important step of all. Logically, this should begin by considering only first-generation properties (u and d quarks), and replicating the same exercise separately for second and third generation quarks.

### III-3. Towards an Alternative Standard Model

We have already made most of the points which need to be made regarding the unified model. When the $A_a^\mu$ fields are replaced by W and Z, in equation 10, we would of course try to use coupling parallel to what is used in electroweak theory (EWT)[10,11,15]. It may seem strange to couple the same W and Z fields that we use in EWT to a completely different set of underlying fields ($\Phi$ and U), but these fields are already coupled to different generations of lepton in EWT without problems. The mathematical analysis should be very similar to what is needed for the steppingstone model, but more complicated.

One simple possibility for the Lagrangian of a "new standard model" might be:

$$\mathcal{L} = -\tfrac{1}{4}B_{\mu\nu}B^{\mu\nu} - \tfrac{1}{4}G_a^{\mu\nu}G_{a\mu\nu} + \tfrac{1}{2}(D^\mu\phi)_a(D_\mu\phi)_a - (1/e^2)V_1(e^2\phi^2) + (D^\mu U)^\dagger(D_\mu U) - (1/e^2)V_2$$
(16)

where:

$$B^{\mu\nu} \equiv \partial^\nu B^\mu - \partial^\mu B^\nu \tag{17}$$
$$G_a^{\mu\nu} \equiv \partial^\mu W_a^\nu - \partial^\nu W_a^\mu + e\epsilon_{abc}W_b^\mu W_c^\nu \tag{18}$$
$$(D^\mu\phi)_a \equiv \partial^\mu \phi_a + e\epsilon_{abc}W_b^\mu \phi_c + e'B^\mu \phi_a \tag{19}$$
$$(D^\mu U) = \partial^\mu U - ie(\tau^a/2)W_a^\mu - ie'B^\mu U \tag{20}$$

and where $\phi^2$, $V_1$ and $V_2$ are exactly as in equations 10. Could it be that the solitons of this simple system are enough to reproduce the quarks and leptons of the standard model (with or without charge rotation) – and give us a finite version of the whole thing, able to explain the new empirical results which the present version of QCD cannot?

We cannot rule out this possibility as yet, because we do not yet know enough about the stable soliton solutions even of equation 1, let alone of this system. Equation 16 does have some of the right properties, which emerge from our discussion of the empirical situation. For example, the masses of the W and Z fields fall out directly and correctly from the usual Higgs mechanism[15]. The $\phi^a$ field itself plays the role of Higgs field here. (We do not need to know the soliton properties for this calculation, because the W and Z are bosons, and do not need to be modeled as solitons.) The $\epsilon_{abc}$ term provides the chirality we need to match electroweak theory, but we do need to know the soliton properties in order to verify that the solitons corresponding to leptons do couple as expected to W and Z. Because the soliton masses all depend on coupling to W, Z and A, we would expect them to reflect powers of α, in line with MacGregor's finding.

Alternatively, equation 2 may be insufficient but close. Perhaps it would be enough to add a new symmetry breaking assumption, or perhaps equation 16 would be enough for everything except neutrinos and CP violations (which are outside of today's standard model as well). We need to understand this system better mathematically before we will be able to "debug" it. Still, for a few moments, it is intriguing to wonder whether equation 16 might already be enough to explain everything we have ever seen in support of the standard model of physics.

Looking beyond this specific model field theory -- our goal here is to represent the electrons, muons and τ leptons as well as the quarks as solitons or as bound mixtures of solitons. Generations themselves should be explained. But this leads to a question: should we represent the electron as a bound system of dyons, or should we assume "charge rotation," in which electric charge as we know it actually corresponds to a mix of what is now viewed as magnetic and electric charge in the HtJR model? The second alternative strikes us as more plausible than the first, but there may be other alternatives as well, or ways to make the second alternative cleaner. In particular, we note that the soliton solutions of the HtJR solution are not fully known, let alone those of the extended HtJR system. The known dyon solutions[21,22] of the HtJR system are basically just U-augmented versions of known monopole solutions to the Georgi/Glashow system. But what if we had started from the known *dyon* (bosonic) solutions[56] of the Georgi/Glashow system, and augmented them? Are such particles already predicted by equations 10 or 16, but simply beyond our limited present knowledge of these systems?

The neutrino presents another interesting puzzle here. It would be rather amusing if we reached a point where the main remaining puzzle in physics were – why don't neutrinos have infinite mass? For the moment, we would propose postponing that question, while other work is done. For the long-term, we would consider at least one very heretical possibility: perhaps the neutrino "lives" between unusual boundary conditions (between its creation and annihilation) such that it can carry away half-integral spin, without being a soliton and without obeying Pauli statistics; perhaps its miniscule "mass" is basically just a field effect, like the masses of the W mesons. But again, there are more urgent issues before us, and much larger experimental numbers crying out for some kind of explanation. If equation 16 or a "debugged" version of equation 16 does happen to have neutrino-like solutions, we may not need to look further in any case.

## III-4. A Possible Approach to Grand Unification

Equation 16 still does not incorporate all of the suggestions in Schwinger's original paper[1]. For example, Schwinger suggested that there should also be some symmetry between the known charged vector bosons W carrying electric charge, and vector bosons carrying magnetic charge. We do not easily see the latter, because the strong power of the magnetic coupling creates a kind of emergent symmetry breaking in what we see at the macroscopic level; however, he argued, they may play a crucial role in explaining CP violations and approximate CP symmetry – and perhaps in the overall menu of particles we observe.

Because the charge of the $W^{\pm}$ bosons basically results from a nonAbelian coupling, the most obvious way to implement Schwinger's suggestion here is to consider a new Lagrangian like:

$$\mathscr{L}_S = -\tfrac{1}{4}B_{a\mu\nu}B_a^{\mu\nu} - \tfrac{1}{4}G_a^{\mu\nu}G_{a\mu\nu} + \tfrac{1}{2}(D^{\mu}\phi)_{ab}(D_{\mu}\phi)_{ab} - V_1(\phi) + (D^{\mu}U)^{\dagger}(D_{\mu}U) - V_2(U^{\dagger}U, \phi) \quad (21)$$

where the field $\phi$ is now an isotopic matrix $\phi_{ab}$ (not assumed to be symmetric or antisymmetric), and where:

$$B_a^{\mu\nu} \equiv \partial^{\mu}Z_a^{\nu} - \partial^{\nu}Z_a^{\mu} + e\epsilon_{abc}Z_b^{\mu}Z_c^{\nu} \quad (22)$$
$$(D^{\mu}\phi)_{ab} \equiv \partial^{\mu}\phi_{ab} + e\epsilon_{acd}W_c^{\mu}\phi_{db} \pm e'\epsilon_{bcd}Z_c^{\mu}\phi_{ad} \quad (23)$$
$$(D^{\mu}U) = \partial^{\mu}U - ie(\tau^a/2)W_a^{\mu}U \pm ie'(\tau^a/2)Z_a^{\mu}U \quad (24)$$

Of course, the choice of the Higgs terms $V_1$ and $V_2$ is important in reproducing not only the solitons, but the general asymmetry we see between electric and magnetic charge. In this theory, the usual scalar neutral Z field is actually just one member of a triplet, in which the other two mesons, being magnetically charged, are much harder to see.

Equation 21 exhibits the symmetry that Schwinger discussed, between electricity and magnetism. But equations 5 and 6 suggest a more far-reaching symmetry. In fact, symmetry under charge *rotation* may be crucial to matching up the theoretical predictions with the empirical data, in the end. Thus we may postulate an alternative version of $\mathscr{L}_S$, in which $\phi_a$ is an SU(3) octet, and in which the W and Z mesons are also part of an octet.

This would be more elegant than equation 21; in the spirit of supersymmetry theory, we might say "therefore it must be true." But we would prefer to adhere to the traditional scientific method, which does not make such commitments until and unless empirical data guides us to a choice.

For a true grand unification, we must of course suggest at least one possibility for how to integrate either of these possibilities with gravity and how to quantize the resulting theory. One such possibility is to use the Lagrangian:

$$\mathscr{L} = (-g)^{\frac{1}{2}} (R - 2\kappa \mathscr{L}_S - S_\phi^{ab} \phi^{ab} - S_U^\dagger U) \qquad (25)$$

where R is the Ricci curvature scalar, where g is the determinant of the metric tensor, where we are following the recipe of Carmelli[57] for metrifying equation 21, and where the final two terms are stochastic terms used to quantize the theory. More precisely, this Lagrangian is a nonautonomous Lagrangian, in which the stochastic source terms $S_\phi$ and $S_U$ act as exogenous variables. They represent time-symmetric Gaussian white noise processes, such that the resulting dynamical system is a mixed forwards-backwards stochastic differential equation[58]. It is important that stochastic differential equations can be well-defined mathematically, without any need to consider or fulfill axioms which refer to Fock-Hilbert space. The variance matrices of the noise terms are parameters of the model; if they are small enough (and not applied to the free boson fields), the net effect is to introduce a small Brownian motion effect in the motion of the solitons, just enough to match the usual statistics in canonical QFT[4]; for the standard model, which is a quasilinear system, the effect is essentially the same as the stochastic perturbation of particle paths predicted by the path integral formulation of quantum theory. Even though this model is stochastic, it is in fact a local realistic model – perhaps the first to be proposed which has any hope of reproducing those predictions of the standard model which have been verified experimentally. (See a summary[4] of our previous work for a discussion of how such a thing might be possible, despite the conventional wisdom which says that it cannot be.)

The source terms in equation 25 would be much smaller than those usually assumed in discussions of vacuum energy or Casimir effects. Nevertheless, they would predict effects more or less equivalent to the cosmological constant term in traditional gravitational theory. In January, 2007, a group in Denmark led by Jesper Sollerman and Tamara Davis was able to estimate a nonzero cosmological constant based on the best data available at that time. An early partial account of their work is available on the web[59]. Based on current estimates of the Hubble constant, their measurement works out to about $7*10^{-30}$ grams of gravitation source per cubic centimeter of space[60].

How could we probe the presence of small $S_\phi$ and $S_U$ effects which are predicted to exist even in the vacuum of outer space? Our only way to probe large volumes of deep space at present is to observe what happens to light and to neutrinos traveling long distances. But because $S_\phi$ and $S_U$ do not interact directly with electromagnetism, we would expect only a very small effect even over distances of light-years. We might expect a very slight loss of energy in each photon, due to thermodynamic effects, and a slight broadening of event signatures in space and in time, but we have not yet computed the details. With neutrinos one might expect larger effects.

Equation 25 would lead to other differences which are testable in principle, compared with the theories of quantum gravity most popular today. For example, it would not predict such a rapid evaporation of black holes as predicted by the Hawkings model. However, even if we could find a way to perform that decisive test, it would be more prudent (and technically easier) to perform it in space.

# APPENDIX:
## PRELIMINARY EXPLORATION OF HtJR STATISTICS

The theoretical research strategy proposed in this paper asks us to start out with a deeper exploration of the HtJR model field theory (equations 10).

The HtJR papers[21,22] and the analysis by Goldhaber[53] provide the most concrete, rigorous analysis of "bosonization" so far in 3+1 dimensions. But at the end of the day, why do we expect such solitons to look like the usual spin-½ fermion, even in the limit where their radius r→0? Many physicists would expect that the usual spin-statistics theorem[8] is enough to guarantee this, but the original discussion by Streater and Wightman[8] stresses that they *assume* that the universe offers us only two choices – bosonic statistics or fermionic statistics. That choice is what we actually see empirically, at a certain scale of length, but it leaves open many important questions about the mathematics which lead to what we see.

Here are two of those questions.

To begin with, consider the family of *classical* solitons of equations 10 with spin ±½ defined by a set of parameters {e(r),…,λ(r)} as discussed in section III-2. Consider the quantized version of these solitons, defined by the quantum corrections described by Rajaraman[13].

First question: assume (*tentatively*) that we can find a family of parameters {e(r),…,λ(r)} such that when r→0: (1) the mass-energy of each soliton goes to some value m>0; (2) the spin remains at ±½; (3) $|\Phi^2(\underline{x})-F^2|$→0 and $U^2(\underline{x})$→0 for any point $\underline{x} \ne \underline{0}$. Does there exist a (dual-valued) mapping from the bosonic wave function of a system of such solitons and the accompanying A fields into the usual kind of Fermi/Bose wave function ψ, and a Fermi/Bose Hamiltonian H, such that the energy of the full wave function always → <ψ|H|ψ> as r→0? I would call this proposition the "A hypothesis."

Based on physical arguments from Louis de Broglie[61], I originally believed that the A hypothesis would probably not be true. However, a careful re-examination suggests that it work out after all. The fermionic bound states described in section II-4 involve a *much lower energy and frequency* (and a longer wave-length) than that of the original "bare" bosonic particles; thus there would be no difficulty in explaining classical experiments like Stern-Gerlach.

Second question: assume that we can find a similar family of parameters, where $U^2(\underline{x}, r) \to U^2(\underline{x}, 0)$, a nonzero limit. Can a suitable Bose to Fermi/Bose mapping exist in *this* case? I called this the proposition the "U hypothesis' in my earlier thinking.

Here is some of the physical intuition behind the U hypothesis.

Louis DeBroglie[61] proposed a realistic physical picture of the electron as a combination of three main elements – a nonlinear "core", a "pilot wave" and an electromagnetic field. In equations 10, the Φ field for |$\underline{x}$|<r (with accompanying values of U and A) meets the requirements for such a nonlinear "core," while the U field outside that core meets the requirements for a "pilot wave." The values for U near the boundary of the core provide the boundary conditions which fixe the values of U, as a "linear wave," outside that core.

In the HtJR description, U is a simple two-spinor; however, if we also allow for two possibilities for the topological charge ("clockwise" and "counterclockwise" Φ), this maps in a natural way into a Dirac 4-spinor field for each possible soliton state. Furthermore, the $U^{\dagger}U( )$ interactions should imply a kind of conserved "lepton number" here, exactly like lepton number conservation which results from the $\psi^{\dagger}\psi( )$ interactions in QED. Thus the space of bosonic wave functions for U is naturally decomposed into disconnected eigenspaces of "lepton number". We may define a mapping from the U wave function of a spin-½ soliton by simply projecting it into the eigenspace which has the right lepton number.

Then, for two such solitons at a distance (noninteracting), consider what we see (when internal details of each soliton become too fine for us to see) when we rotate the system

180 degrees around its center of gravity; the fermionic statistics work, but any bosonic components would be smeared away to zero.

Finally, as we allow interaction, by bringing the solitons to a finite distance from each other (but still "much bigger than r"), our lepton number result remains valid, and allows the approximation to still work.

This deals with "physical" solitons. For virtual pairs – the vacuum polarization correction to the A field might seem to be a problem at first, since it depends on the true, fine-scale propagator of U, which is different from the Dirac propagator. It only yields a fourth-order term to be integrated. However, in QED, renormalization is used to eliminate the terms with lower-order denominators; with correct choices of parameters, there is reason to hope that the correction to A would be the same in both cases.

In re-examining this situation, it seems clear that the "A" hypothesis provides a different way to address De Broglie's concerns, while leading to much simpler mathematics. "Bosonization" occurs by simply imposing the requirement that the angular momentum of the wave function in collective coordinates must match the angular momentum of the underlying bound system. Vacuum polarization terms come out right because the terms involving naked high-energy bosons have very low probability amplitude. The "pilot wave" is simply the wave function in collective coordinates. When we apply classical-quantum equivalence relations to the underlying bosonic wave functions, we see a picture of a "chaotic electron," where deBroglie's "core" skitters between the two slits in a two-slit interference experiment. In that picture, the ordinary long wavelength of the electron is somewhat analogous to the low frequency acoustic waves which modulate the high frequency carrier, in an AM radio signal.

In summary, a suitable (dual-valued) mapping from {U, A} to {ψ, A} does seem to be available. However, a complete mapping should also account for the Φ field. But as r→0, all we see of Φ (as predicted by the bosonic field theory) is the topological charge and an effect on the mass and charge which is all concentrated at a single point, like δ(**x**). The effect of this is really quite beautiful. It provides a kind of *physical explanation* for the δ(**x**) renormalization terms now added to the mass and charge of the electron, in today's QED. We are mapping from a family of f*inite* bosonic field theories into the *corrected, renormalized* version of the usual Fermi/Bose field theory!

Perhaps, then, a "topological charge" of +1 in the HtJR model actually refers to a kind of linear mix ($e_e$ and $g_e$) of what we call electric and magnetic charge. Perhaps the U and A fields in the core of the electron somehow shield us from the magnetic part of this charge, if it is nonzero.

In the end, this reasoning seems to be point towards a finite, bosonic version of QED, rather than a revision of QCD – but nailing down the mathematics of this simple model is an essential starting point. There are two major theoretical tasks ahead of us: (1) to really prove or debug/prove these basic properties of spin and statistics, and address other aspects of the HtJR Model, as described in section II-3; and (2) to explore various possible extensions of the HtJR model, similar in flavor to those discussed in sections III-3 and III-4, and guided by the empirical results summarized in Section II.